\documentclass{elsart}

\usepackage{amssymb}
\usepackage{graphicx}

\begin{document}

\begin{frontmatter}



\title{A new concept for streamer quenching in resistive plate chambers} 


\author[label0]{A.~Calcaterra}
\author[label0]{R.~de~Sangro}
\author[label1,label0]{G.~Mannocchi}
\author[label0]{P.~Patteri}
\author[label0]{P.~Picchi}
\author[label0]{M.~Piccolo}
\author[label2]{N.~Redaelli}
\author[label2]{T.~Tabarelli de Fatis \corauthref{cor1}}
\author[label1,label3]{G.C.~Trinchero}

\address[label0]{Laboratori Nazionali di Frascati, INFN, \\ 
                 Via E.~Fermi 40, I-00044 Frascati (RM), Italy}
\address[label1]{Istituto Nazionale di Astrofisica, IFSI Sezione di Torino, \\ 
                 C.so Fiume 4, 10133 Torino, Italy}
\address[label2]{Universit\`a di Milano-Bicocca and 
                 INFN, Sezione di Milano, \\
                 Piazza della Scienza 3, I-20126, Milano, Italy}
\address[label3]{INFN, Sezione di Torino, \\
	         Via P. Giuria 1, I-10125, Torino, Italy} 
\corauth[cor1]{Corresponding author: tommaso.tabarelli@mib.infn.it}

\begin{abstract}

In this paper we propose a new concept for streamer quenching in
Resistive Plate Chambers (RPCs). In our approach, the multiplication 
process is quenched by the appropriate design of a mechanical 
structure inserted between the two resistive electrodes. 
We show that stable performance is achieved with binary gas mixtures 
based on argon and a small fraction of isobutane. Fluorocarbons, deemed 
responsible for the degradation of the electrode inner surface of RPC
detectors, are thus fully eliminated from the gas mixture. This 
design {also resulted} in a simplified assembly
procedure. Preliminary results obtained with a few prototypes of 
``Mechanically Quenched RPCs'' and some prospects for future
developments are discussed. 

\end{abstract}

\end{frontmatter}

\section{Introduction}
\label{intro} 

Resistive Plate Chambers (RPCs), gaseous detectors with parallel
resistive plates \cite{San81,San88}, can provide both good spatial
information and time resolution  in detecting charged particles over
large areas at low cost.

Besides the traditional use of RPCs as a low cost coarse tracker, to
identify muons and provide an estimate of their momenta, new
prospects for their applications as the active part of a digital
hadron calorimeter at the International Linear Collider experiment
\cite{digiCa} and for X- or gamma-source imaging  \cite{Fran01} are
emerging. These are spurring again the R\&D program toward a reliable
detector, suited for large scale application in modern experiments,
where industrial standards for mass  production with quality control
and assurance and stable performance over the long term are
compulsory.

RPCs can be operated in avalanche mode and in streamer mode. The
former has an advantage in terms of a higher rate capability, while
the latter carries an advantage in terms of larger signals (order of
100 pC).  A typical gas mixture for RPC operated in streamer mode
consists of argon (Ar), butane (C$_4$H$_{10}$) and a non-flammable
fluorocarbon  (freon).  The ternary mixtures have been proved to
provide stable operation with high efficiency to detect minimum
ionising particles.  Operations without freon have also been studied
by  many groups and various types of freon-less gas mixtures or
mixtures with very low freon concentration have been reported
\cite{many01,many02,many03,many04,many05,many06}.  An interesting
alternative to freon is the use of SF$_6$, which has a very high
electron affinity and enables for reasonable streamer quenching at
concentrations as small as 1\% \cite{Abe00}. These attempts, however,
never resulted in a large scale  application of RPCs, where
non-flammable and low cost gas mixtures are required. Thus, gas
mixtures with a large fraction of an environmentally friendly freon,
the tetrafluoroethane (CH$_2$FCF$_3$) known as HFC-134a , are commonly
used \cite{BaBar,Belle}.  It is however desirable not to use these
gases in the future, as they  are being recognised as the main source
of RPC instability in the long  term. Indeed, by decomposition of the
tetrafluoroethane, under electrical discharges, hydrofluoric acid (HF)
can be formed, which is a very aggressive one and is expected to have
a main role in damaging the RPC inner surface \cite{San04}. In
particular, the corrosive effect of HF in RPC detectors with glass
electrodes is a fully fledged phenomenon \cite{Japan,Capire}. HF acid
has long been known in glass and silicon industry for its etching
properties, indeed.

In this paper we propose a new concept for streamer quenching, based
on a completely different approach. We argue that the role of the
freon gas can be fully replaced by the appropriate design of a
mechanical structure inserted between the two resistive electrodes. In
practice this is achieved by introducing a honeycomb structure 
{opaque to UV photons} 
between the two resistive electrodes to keep the streamer confined in 
one single cell.  There are several technical problems that need to be
addressed: 
the honeycomb material should be an excellent insulator; the cell
dimensions must be tuned to the streamer area; the cell walls must be
thin enough to avoid a geometrical loss of efficiency; a way to
let the gas flow through the chamber need to be found. Yet a material
almost ideal for this kind of application is commercially available.

In the following, after the description of the mechanical realisation
of the first prototypes and of the experimental setup adopted for a
preliminary characterisation of the detectors, we show that stable
performance is achieved with binary gas mixes based on argon and a
small fraction of isobutane. 

\section{Detector description and measurement setup}
\label{detector}

The basic element of our ``Mechanically Quenched'' RPCs is a  ECA-I
core honeycomb structure\footnote{ECA-I is a product name in the
catalogue of Euro-Composites, Zone Industrielle, L--6401, Echternach,
Luxembourg.}, made by an aramid fibre paper coated with phenolic
resin. This material has a high strength-to-weight ratio, is
electrically insulating, chemically, corrosion and flame
resistant. Honeycomb sheets are available with hexagonal cell sizes
from 3.2~mm to 19.2~mm diameter and minimum thickness of 2~mm. The
cell walls are less than 50~$\mu$m thick for the lowest cell sizes,
giving a geometric fill-factor, defined as the ratio of the
potentially active area to the total area, in excess of 90\% for
3.2~mm wide hexagonal cells.
Moreover, a direct measurement of the light transmission through 
the aramid fibre paper have shown that the cell walls are opaque 
to UV photons up to 360~nm and transmit only about  10\% of the 
visible light.

\begin{figure*}[tb]
\centering
\includegraphics[width=134mm]{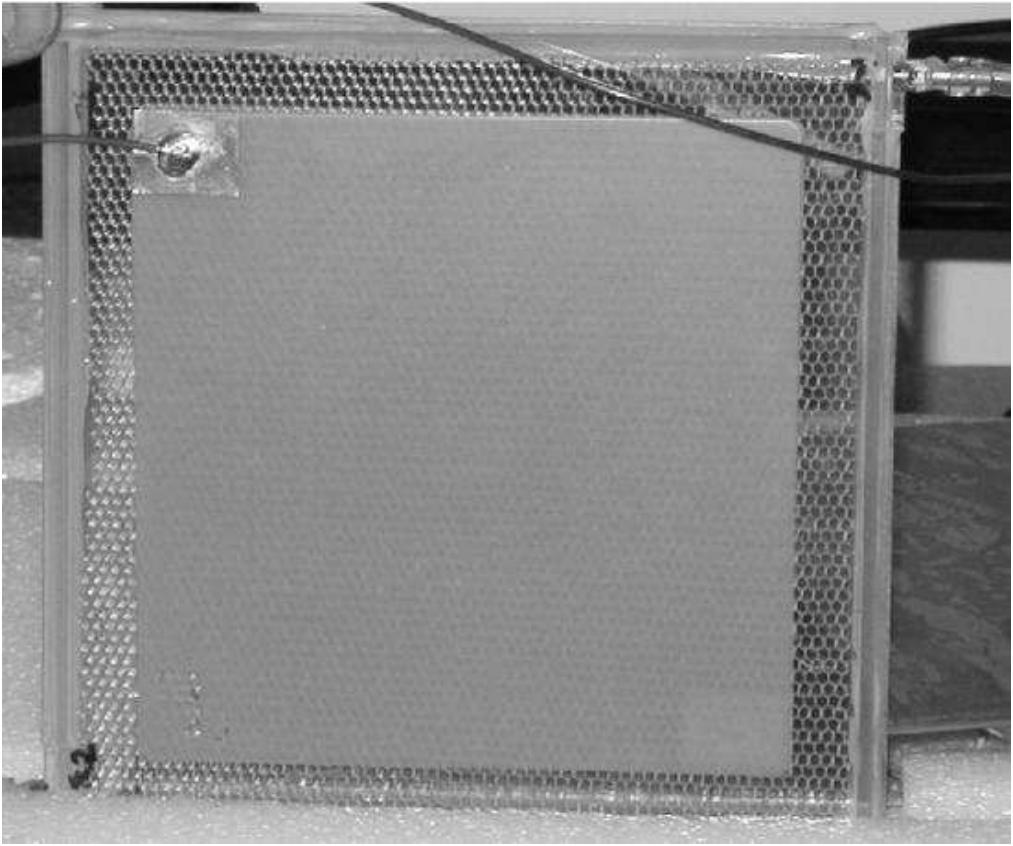}
\caption{One RPC prototype with the honeycomb structure inserted
  between the glass electrodes. The silk screen printed coating and
  the electrical contact for supplying the high voltage are visible.}
\label{chamberpict}
\end{figure*}

We constructed test RPCs 20$\times$20~cm$^2$ in size using resistive
electrodes of 2~mm thick float glasses with $4\times10^{12}~\Omega$cm
bulk resistivity at room temperature. One honeycomb foil  2~mm thick
with 3.2~mm hexagonal cells was inserted between the electrodes. An
external frame of polycarbonate, to which the glass electrodes were
glued, guaranteed the gas tightness of the chamber (see Figure
\ref{chamberpict}). 
This assembly procedure was simpler than for standard RPCs, as the
honeycomb sheet guaranteed a constant gap between the electrodes
with no need for specific spacers.
Prototypes with 3~mm thick glasses were also built
and tested, but the high bulk resistivity of the glass batch used for
these chambers ($>10^{13}~\Omega$cm) resulted in poor performance. The
high voltage was supplied to the outside surface of the glasses by a
thin pre-coated paint (silk screen printed) covering an area of
18$\times$18~cm$^2$ with 200-400 k$\Omega/\Box$ resistivity. The
signal was readout on external pickup copper electrodes, covered with
a PET sheet for electrical insulation from the chamber. %
In most of the measurements, one single copper readout pad, covering 
the entire detector surface was used, thus no position  information
was available. Both analog and digital readout of the signal was
performed, to measure the streamer charge, the chamber noise (single
count rate) and the time resolution. Digital information was obtained
after signal discrimination with -30~mV threshold.

The performance of the RPC prototypes was tested with cosmic rays. The
coincidence of two scintillators, placed above and below the chambers,
provided a trigger over an area of about 8$\times$8~cm$^2$ around the
chamber centre. At every trigger, the ADC (LeCroy 2249A) and TDC
(LeCroy 2228A) information were readout, as well as the content of a
scaler, which recorded the single rate and the three fold coincidences
of each chamber with the two scintillators. 
For charge measurements, the ADC gate was set to 250 ns, well in
excess of the pulse duration, and the pulses were attenuated by 12 dB 
at the ADC input to match the ADC dynamical range. The start to the 
TDC for time measurements was given by the bottom scintillator, with
one stop for each RPC.

As the honeycomb structure was expected to play the role of HFC-134a
in confining the transverse dimensions of the streamer, 
the RPCs were tested with binary Ar/isobutane
non-flammable mixtures. The isobutane concentration was varied from
2\% to 8\% by means of rotameters. The range of operations of the
rotameters available for this test implied a relatively high gas flow
rate of around 50~cc/min in total, corresponding to the change of one
chamber volume every 2~min, approximately. The gas inlet and outlet
were  located on the external frame of the RPCs (see
Fig. \ref{chamberpict}) and no special care was taken in the design of
the chamber to ease the gas flow through the honeycomb. This might
require further consideration in the future. 
It is an experimental fact that, 
at the flow rates used in these tests, there is enough room
for the gas to flow between the inner surface of the electrodes and 
the honeycomb sheet\footnote{The thickness tolerance of the honeycomb 
sheets is about 0.2 mm.}.

\section{Detector performance}
\label{performance} 

A preliminary characterisation of the detector performance was
accomplished by studying the detection efficiency to cosmic rays, the
signal pulse shape and the time resolution. 
The results reported hereafter describe the performance of the two RPC
prototypes assembled with 2~mm thick glass electrodes.  
After assembly, the RPC prototypes were fluxed, at the above
mentioned rate, with an Ar/isobutane 97/3 mixture and conditioned at 
5 kV for about one week, until stable efficiency and single count rate
were achieved. 

\subsection{Efficiency and single count rate}

\begin{figure*}[htb]
\centering
\includegraphics[width=135mm]{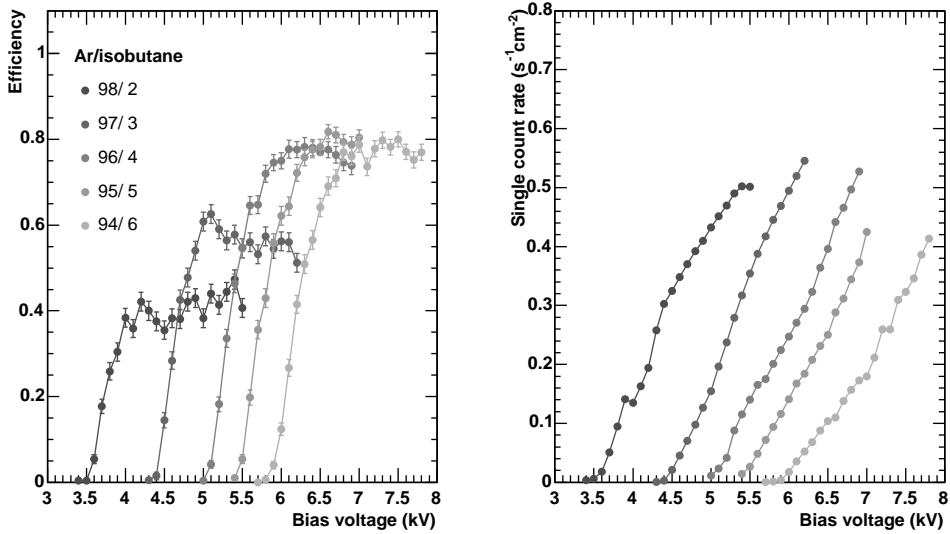}
\caption{Efficiency (left) and single count rates (right) of the RPC
prototypes exposed to cosmic rays.} 
\label{effPla}
\end{figure*}

The efficiency plateau and the single count rate curves of the RPC 
for several Ar/isobutane mixtures are shown in Fig. \ref{effPla}. 
The curves shift toward higher voltages when increasing the
fraction of isobutane. At small isobutane fractions ($<$4\%) the chamber was 
not efficient, as secondary streamers were not sufficiently
quenched. In the limit of pure Ar, not reported in the figure, the
chamber was completely inefficient, because of long after-pulses.  
Isobutane fractions larger than 4\% enabled stable RPC operation, 
although with an inefficiency somewhat larger than expected 
from geometrical considerations. The best performance was observed
with a 95/5 mixture, showing a plateau efficiency of about 80\% for a
supplied high voltage of about 6.5~kV. The efficiency loss may be due
to several reasons that require further investigation. The extension
of the inefficient area around the honeycomb has to be investigated in
detail: the field quality there might be such that the efficiency loss
region is wider than the aramid structure itself. Another source of
inefficiency may be related to the low rate capability of the glass 
electrodes and the relatively high count rate of spontaneous
streamers observed. 

The single count rate as a function of the bias supply was also
measured (see Fig. \ref{effPla}). We observed a fast rise of the
single rate with no evidence of any plateau. This is an indication
that the single count rate was dominated by spontaneous (spurious) 
streamers and not by physics events. At the voltage supply
corresponding to the beginning of the efficiency plateau, a single count
rate of about 0.2~Hz/cm$^2$ was observed. This was four to five times 
larger than the single count rate observed with standard glass
chambers without the the honeycomb insert and operated with a standard
Ar/isobutane/HFC-134a mixture 30/8/62 \cite{Capire}. 

The dark current of the detector showed the same HV dependence of the
single count rate. The contribution to the current due to leakage in
the honeycomb structure is small, as the whole current observed could
be accounted for by the single count rate and the observed charge in
the pulses.  

\subsection{Pulse shape and charge}

\begin{figure*}[htb]
\centering
\includegraphics[width=65mm]{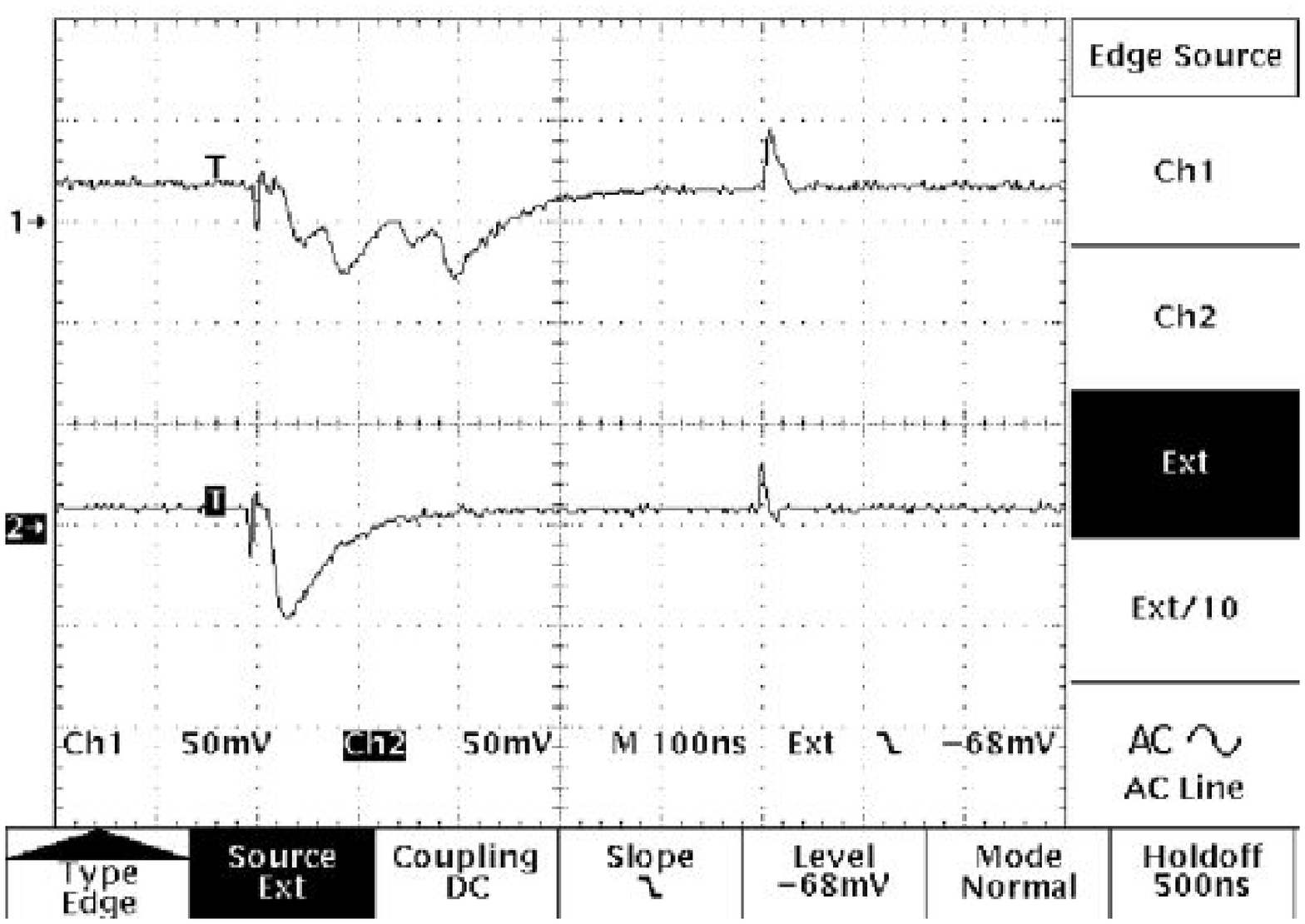}
\includegraphics[width=65mm]{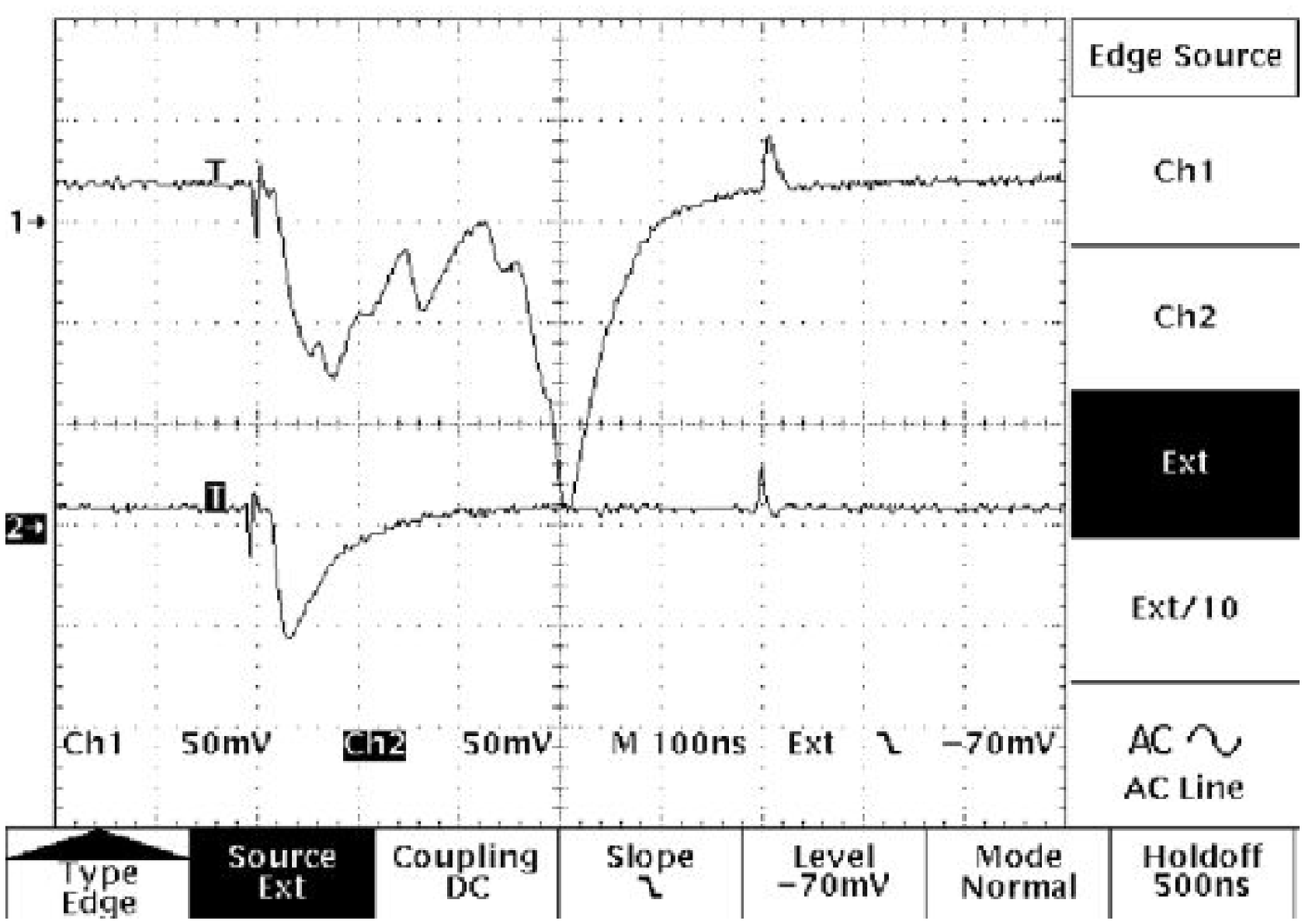}
\caption{Two examples of pulses (left and right frames)
  for Ar/isobutane mixture 95/5 at 6.5 kV in
  RPCs without (top trace) and with (bottom trace) the honeycomb 
  insert upon an external trigger given by a pair of scintillator 
  counters. The signal was terminated on a 25~$\Omega$ resistance 
  and an attenuation of 12 dB was applied.}
\label{twoPulses}
\end{figure*}

Clean single pulses were observed in the test RPCs with honeycomb
inserts (see Fig. \ref{twoPulses}), with high pulse heights
of typically 300 mV on the 25~$\Omega$ load resistance. The fraction
of the signals with multi-pulse is observed trough an oscilloscope to
be less than 10\% for isobutane fractions larger than 4\%. This is at full
variance with the behaviour of a standard (no honeycomb) RPC operated 
under the same conditions (see Fig. \ref{twoPulses}), where multiple 
pulses were in general observed. This testifies of the effectiveness of
the honeycomb structure in keeping the streamer localised within one
single cell and in realising the streamer quenching. On the other
hand, the absence of a fluorocarbon gas made the quenching inefficient
in the standard chamber.

\begin{figure*}[htb]
\centering
\includegraphics[width=125mm]{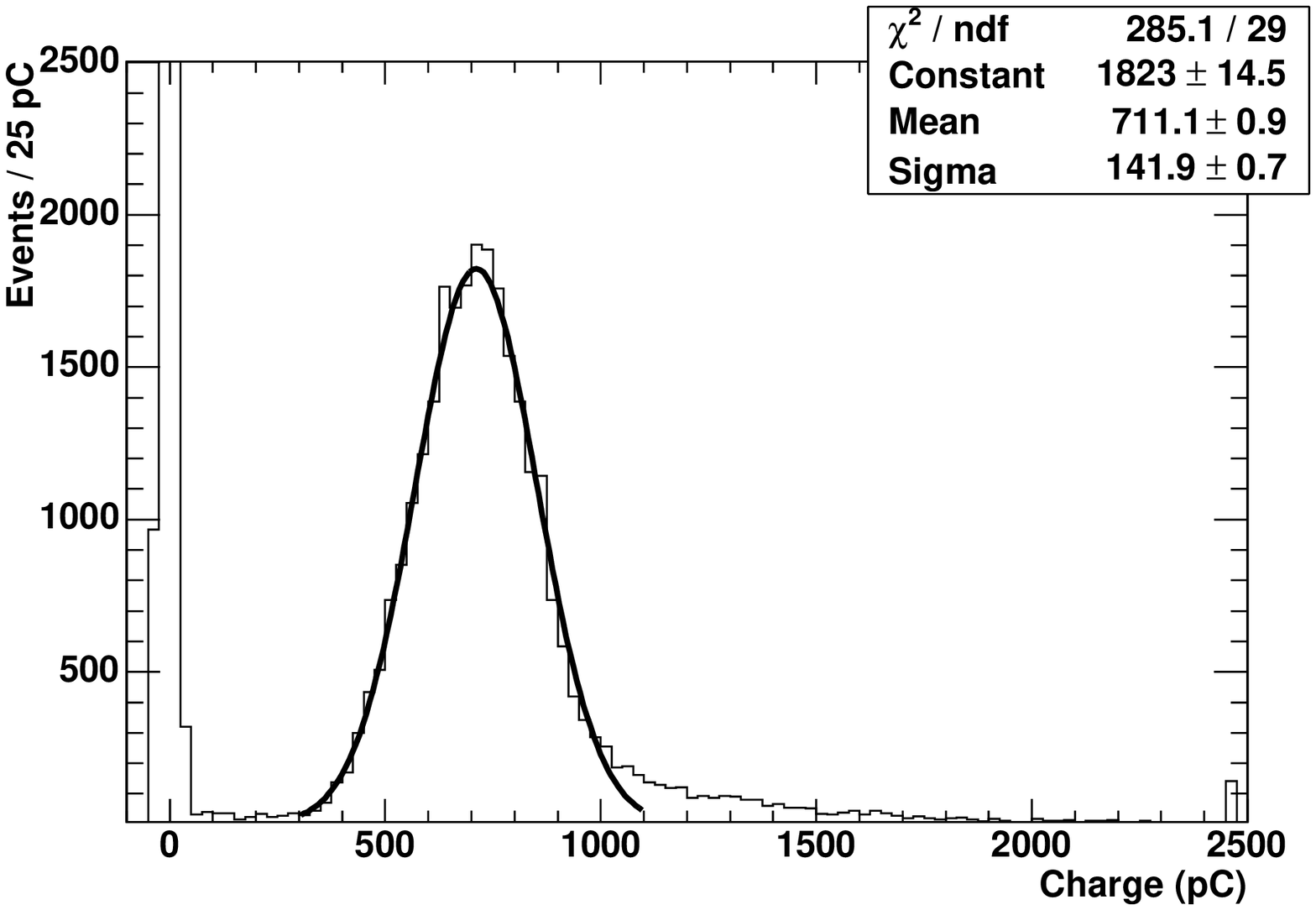}
\includegraphics[width=125mm]{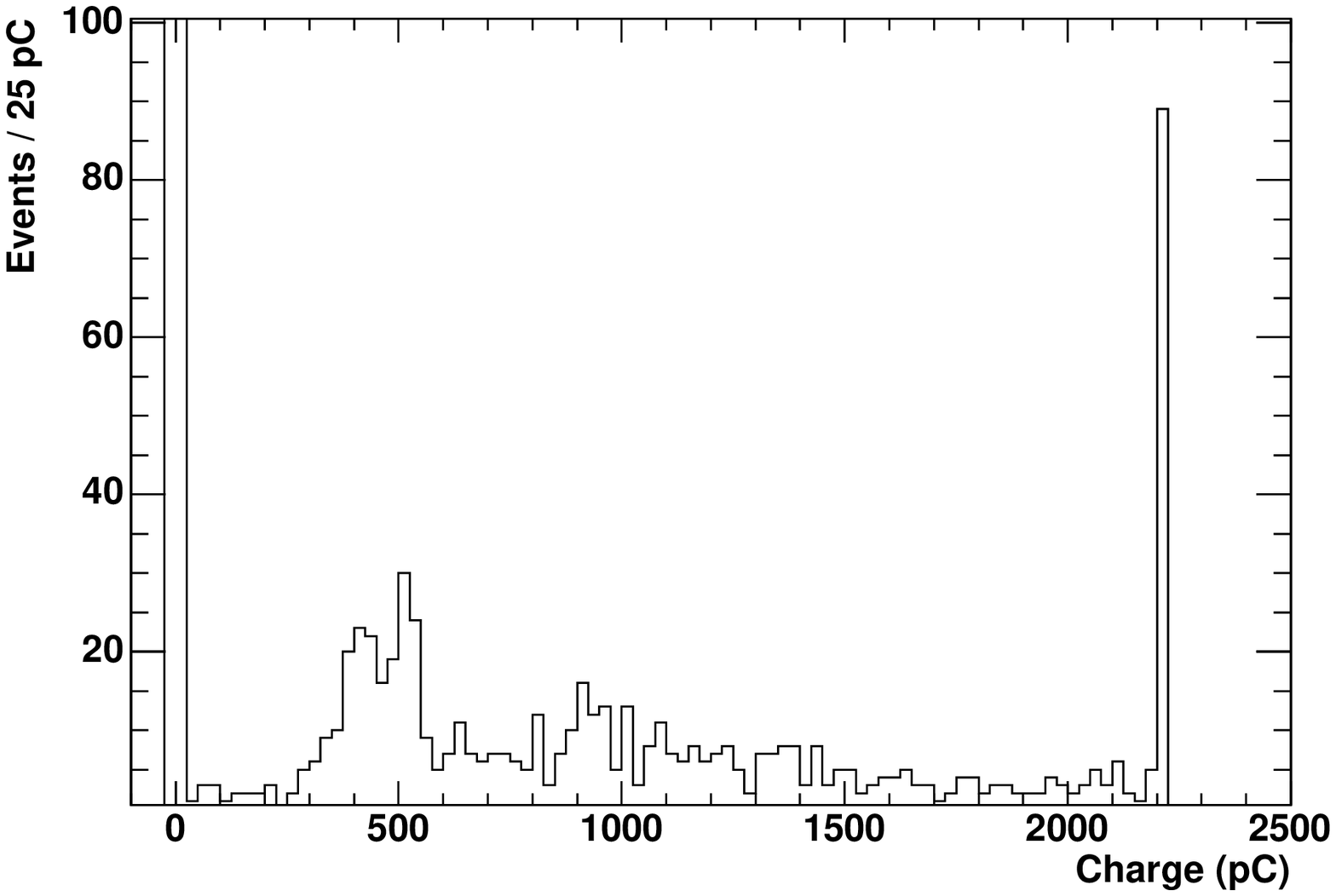}
\caption{Charge distributions at the plateau HV of 6.5 kV with 
  Ar/isobutane 95/5 mixture for the honeycomb RPC (top) and a standard
  RPC (bottom).}
\label{chargeSpectra}
\end{figure*}

These qualitative indications are confirmed by the results shown in
Fig. \ref{chargeSpectra},  where the charge distributions observed in
the honeycomb RPC prototypes and in the standard RPC detector, both
operated at the plateau HV with a Ar/isobutane 95/5 gas mixture, are
compared.  The charge distribution for the honeycomb RPC shows a major
first peak   with a small tail of large pulses. A probability of
multi-pulse signals of about 7\% can be estimated from the total area
of the spectrum above the pedestal and the area of a Gaussian
distribution of about  700 pC mean charge and 140 pC RMS describing
the single pulse peak.  The standard RPC, on the contrary, shows an
indication of the single pulse peak at around 500 pC and a long tail
up to very high signal charges corresponding to multi-pulse
events. The multi-pulse probability is about 70\% with almost half of
the pulses saturating the ADC upper edge at 2.2 nC. 
Due to the inefficiency of the quenching process, the standard RPC
also showed a detection efficiency lower than the honeycomb RPC.

\begin{figure*}[htb]
\centering
\includegraphics[width=125mm]{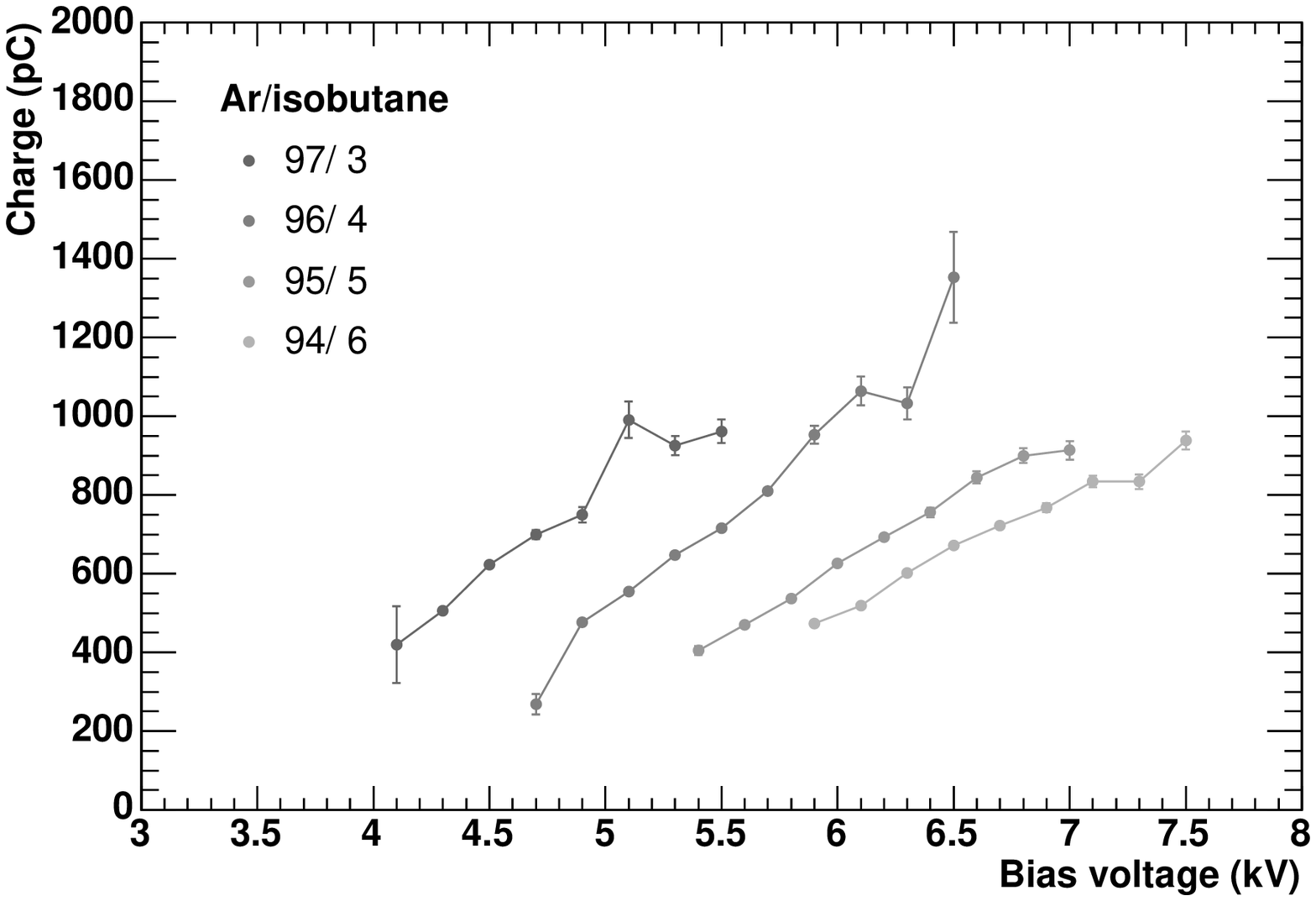}
\caption{Peak charge as a function of the supplied high voltage for
  different gas mixtures.}
\label{chargeHVdependence}
\end{figure*}

The HV dependence of the charge distribution measured showed almost
the same shape for the honeycomb structure, with only a small increase
in the multi-pulse probability at the higher voltages. The peak charge
induced on the readout pad versus the high voltage for each gas
mixture is shown in Fig. \ref{chargeHVdependence}. The charge
exhibits an approximately linear dependence on the operating HV,
distinctive of 
{space charge dominated avalanche growth.}
The induced charge is only
mildly dependent on the isobutane fraction, but always rather large as
compared to a typical charge of about 100 pC of the streamer mode
operation with freon gas. This is explained qualitatively by the fact 
that the absence of a gas with high electron affinity such as
freon leads to a higher gas gain. 

The streamer charge is expected to increase with the cell dimensions. 
A preliminary indication of this effect was obtained with a test RPC 
detector assembled with a honeycomb sheet modified to obtain larger 
elementary cells. The charge dependence on the cell size needs to be 
further investigated and systematic measurements with honeycomb sheets 
of different cell sizes are being planned.

\subsection{Time information}

\begin{figure*}[htb]
\centering
\includegraphics[width=125mm]{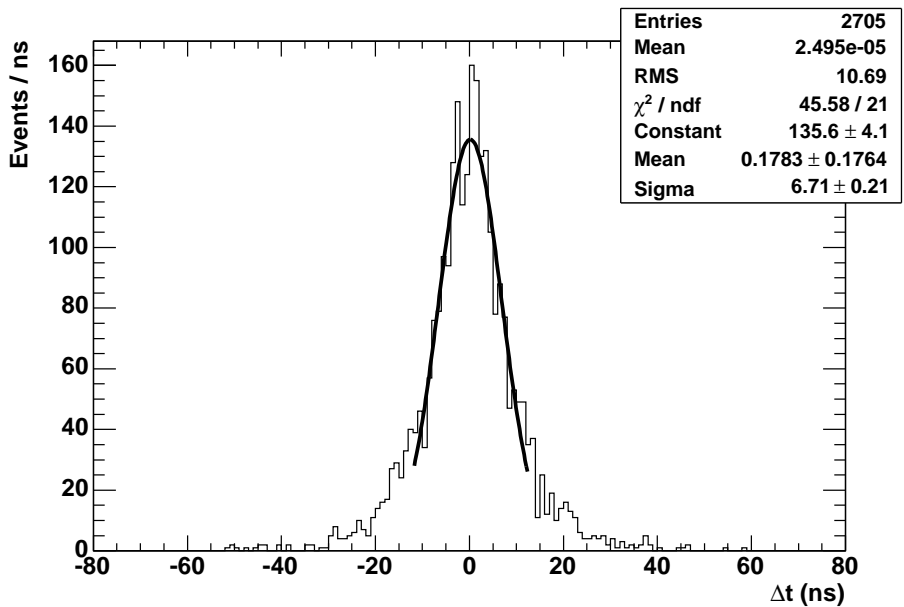}
\caption{Time difference in the response of two test RPCs operated at
  6.5~kV with an Ar/isobutane 95/5 mixture.} 
\label{timeResponse}
\end{figure*}

A preliminary measurement of the time resolution of the test RPCs can
be seen in Fig. \ref{timeResponse}, where the distribution of the time
difference between two chambers, operated at 6.5~kV with an
Ar/isobutane 95/5 gas mixture, is plotted. Assuming that the
resolution of the two chambers was identical, a RMS of about 7.5~ns
for each chamber could be estimated. 
{Part} of the large RMS can
be ascribed to the non-Gaussian tails of the distribution.  If a
Gaussian fit to the core of the distribution is performed, a
resolution of 5~ns is obtained. 
This resolution only slightly improved at higher voltages. 

\section{Summary and outlook}
\label{conlcusion} 

A few prototypes of ``Mechanically Quenched'' glass RPCs 
were built and tested. In these detectors, the quenching and
transverse confinement of the streamer was obtained by means of
a honeycomb structure inserted between the two resistive electrodes. 
We have shown that this detector design allowed stable operation
in streamer mode with Ar/isobutane non-flammable gas mixtures. 
The optimal performance was achieved for isobutane fractions around 5\%. 
This represents a new and promising approach to operate RPC detectors
with gas mixtures not containing freon, which is recognised as the
main source of RPC instability in the long term.

The preliminary results reported in this paper are encouraging,
although the chambers did show a somewhat larger inefficiency than
expected from geometrical considerations and the typical single count
rate (0.2 Hz/cm$^2$) was about four times larger than in glass RPCs
with standard Ar/isobutane/freon mixtures.  
{Being essentially a new technology, many aspects require
further investigation. In particular, the optimisation of the 
honeycomb geometry, which affects the detector efficiency and 
the streamer charge, and of the electrode material, which affects the
rate capability and thus the efficiency of a detector with a
relatively high rate of spontaneous streamers, has to be addressed. 
In this prospect, RPC prototypes with different cell dimensions and
with bakelite electrodes are being produced. 
If high efficiencies can be demonstrated, a careful study of the
mechanical design and of the long term behaviour of these chambers
will also be deemed mandatory. } 

\section*{Acknowledgements}

The skillful work of R. Bertoni is warmly acknowledged. We also thank
our students G.~Croci, F.~De Guio and N.~Ketz for their valuable
contributions.

\end{document}